\documentclass[preprint]{emulateapj}
\begin{document}
\title{
Time-dependent force-free pulsar magnetospheres: axisymmetric and oblique rotators
}
\author{Anatoly Spitkovsky\altaffilmark{1}}
\altaffiltext{1}{ Kavli Institute for Particle Astrophysics and Cosmology,
Stanford University; anatoly@slac.stanford.edu}
\subjectheadings{pulsars -- neutron stars -- magnetospheres }
\begin{abstract}

Magnetospheres of many astrophysical objects can be accurately described by the low-inertia (or ``force-free") limit of MHD. We present a new numerical method for solution of equations of force-free relativistic MHD based on the finite-difference time-domain (FDTD) approach with a prescription for handling spontaneous formation of current sheets. We use this method to study the time-dependent evolution of pulsar magnetospheres in both aligned and oblique magnetic geometries. For the aligned rotator we confirm the general properties of the time-independent solution of Contopoulos et al. (1999). For the oblique rotator we present the 3D structure of the magnetosphere and compute, for the first time, the spindown power of pulsars as a function of inclination of the magnetic axis. We find the pulsar spindown luminosity to be $L\approx (\mu^2 \Omega_*^4/c^3) (1+\sin^2\alpha)$ for a star with the dipole moment $\mu$, rotation frequency $\Omega_*$, and magnetic inclination angle $\alpha$. We also discuss the effects of current sheet resistivity and reconnection on the structure and evolution of the magnetosphere. 

\end{abstract}
\maketitle
\section{Introduction}
Many astrophysical objects, including neutron stars and accretion disks, form highly magnetized magnetospheres. Modeling of the structure of such magnetospheres requires solving for the self-consistent behavior of plasma in strong fields, where field energy can dominate the energy in the plasma. This is difficult to do with the standard numerical methods for MHD which are forced to evolve plasma inertial terms even when they are small compared to the field terms. In these cases it is possible to reformulate the problem and instead of solving for the plasma dynamics in strong fields, solve for the dynamics of fields in the presence of conducting plasma. This is the approach of force-free relativistic MHD, or force-free electrodynamics (FFE; see, e.g., Komissarov 2002). In this Letter we present a new numerical method for solving the equations of FFE and apply it to the problem of structure of pulsar magnetospheres. We calculate the shape of the magnetosphere for arbitrary magnetic inclination and determine the pulsar spindown power. 

\section{Numerical method}

The equations of FFE can be derived as the low-inertia limit of MHD (Komissarov 2002). When plasma inertia and temperature are small, the balance of forces on the plasma is $\rho {\bf{E}}+{1\over c}{\bf{j}}\times {\bf{B}}=0$ (``force-free" condition), where $\rho$ and $\bf{j}$ are charge and current densities. When plasma is perfectly conducting and abundant to short out the accelerating electric fields (${\bf E}\cdot {\bf B}=0$), the Maxwell equations in special relativity together with the force-free condition give (Gruzinov 1999; Blandford 2002): 
\begin{eqnarray}
{1\over c} {\partial  {\bf E}\over \partial t}={\bf \nabla} \times {\bf B}&-&{4 \pi \over c} 
{\bf j}, \quad {1\over c} {\partial {\bf B}\over \partial t} =-{\bf \nabla} \times {\bf E}, \label{ff1} \\
{\bf j}={c\over 4 \pi} {\bf \nabla} \cdot {\bf E} {{\bf E} \times {\bf B} \over B^2}&+&
{c\over 4 \pi} {({\bf B}\cdot {\bf \nabla}\times {\bf B}-{\bf E}\cdot {\bf \nabla}\times {\bf E}){\bf{B}}
\over B^2}\label{ff3}
\end{eqnarray}
Eq. (\ref{ff3}) is a prescription for the plasma current that is needed to satisfy the constraints. The two terms have simple meaning: the first is the current perpendicular to fieldlines, given by the advection of charge density with the ${\bf E}\times {\bf B}$ drift velocity, while the second term is the parallel component of the current which reduces to $({\bf \nabla}\times {\bf B})_{\parallel}$ in steady state. The system (\ref{ff1}-\ref{ff3}) is hyperbolic, can be evolved in time, and is much simpler than the full MHD equations.

In order to numerically solve this system we utilize its close relation to Maxwell equations and use the finite-difference time-domain (FDTD) method commonly used in electrical engineering (Yee 1966). Its strengths are very low numerical dissipation and conservation of divergence-free condition to machine accuracy. Electric and magnetic fields are decentered component-wise on an orthogonal grid with electric components defined in the middle of cell edges and magnetic components in the middle of cell faces, allowing to compute the change of flux through cell face from circulation around cell edges. We use third-order Runge-Kutta time integration, maintaining time alignment of the fields. To calculate the current in (\ref{ff3}) we use linear interpolation of the fields to the same locations on the grid (at the $E$-points). The second term in (\ref{ff3}) comes from the perfect conductivity constraint and is cumbersome to calculate numerically as it requires interpolation of both the fields and field derivatives. Instead of using the full eq. (\ref{ff3}) we solve an even simpler system: we advance Maxwell equations with only the perpendicular component of the current, and then after every timestep we subtract the accumulated $E_{\parallel}$ component from the electric field to enforce ${\bf E}\cdot {\bf B}=0$. This is equivalent to replacing the parallel current with a resistive term $\sigma_{\parallel} E_{\parallel}$ (Komissarov 2004), where the parallel conductivity $\sigma_{\parallel}$ is taken to infinity. 

There are two conditions for the validity of the force-free approximation: electromagnetic energy density vastly exceeds plasma pressure and inertia, and the plasma drift velocity is subluminal, so that $E^2\leq B^2$. The system (\ref{ff1}-\ref{ff3}) does not enforce these conditions, and initial force-free configuration may develop regions that violate one or both conditions. When $E$ exceeds $B$, as may happen during the interaction of strong waves, the Alfven speed becomes imaginary and leads to an exponential instability, which physically would be averted by dissipation. Another example is the formation of current sheets, which is a generic feature of force-free evolution. In current sheets magnetic field can reverse and go through zero. This can violate both conditions, which means that the assumption of magnetic dominance is invalid and one has to restore plasma effects such as pressure to sustain the current sheet.  However, in the framework of FFE these effects are not included, so we try to keep current sheets unresolved on the grid, with a jump in fields over one cell. The FDTD method allows to support such sharp discontinuities without spreading them. In order to have the method capture current sheets, we enforce the condition $E\leq B$ after every timestep by resetting the electric field to $E=B$ in the regions where it exceeds $B$. The physics behind this drastic measure is related to dissipation processes that will happen if ${\bf E} \times {\bf B}$ drift approximation is violated and particles accelerate while decoupling from magnetic field lines. This dissipation will restore $E=B$ on plasma timescales; however, this physics remains to be clarified.   We parametrize this physics with prescriptions for the dissipative current $j_{\perp}$ that is added to (\ref{ff3}) when $E>B$. The prescription we use can be formulated as $j_\perp=\sigma_\perp E$ when $E>B$ and $j_\perp=0$ otherwise, where the perpendicular conductivity $\sigma_\perp \to \infty$. Other prescriptions are also possible (Komissarov 2004). 

Our method is not the only way to solve FFE. 
By formulation our method is similar to Komissarov's (2004, 2005), in that we also solve the non-conservative form of the equations with explicit resistive current prescriptions. The limiter procedure for enforcing $E<B$ is similar to what was used in McKinney's conservative scheme (2006a). The main difference, however, is that our FDTD discretization has inherently much lower numerical diffusivity. This is why our method is able to develop and sustain sharp current sheets unlike Komissarov's (2005) scheme, without the need to explicitly null the inflow velocity into current sheets as in McKinney (2006a). In fact, the location of the sheet does not have to be known in advance, which makes the method useful for evolution of complicated initial conditions. 
Below we describe the applications of our code to pulsars. 
 
\begin{figure*} 
\unitlength = 0.0011\textwidth
\hspace{1\unitlength}
\begin{picture}(280,400)(0,20)
\includegraphics[scale=.55]{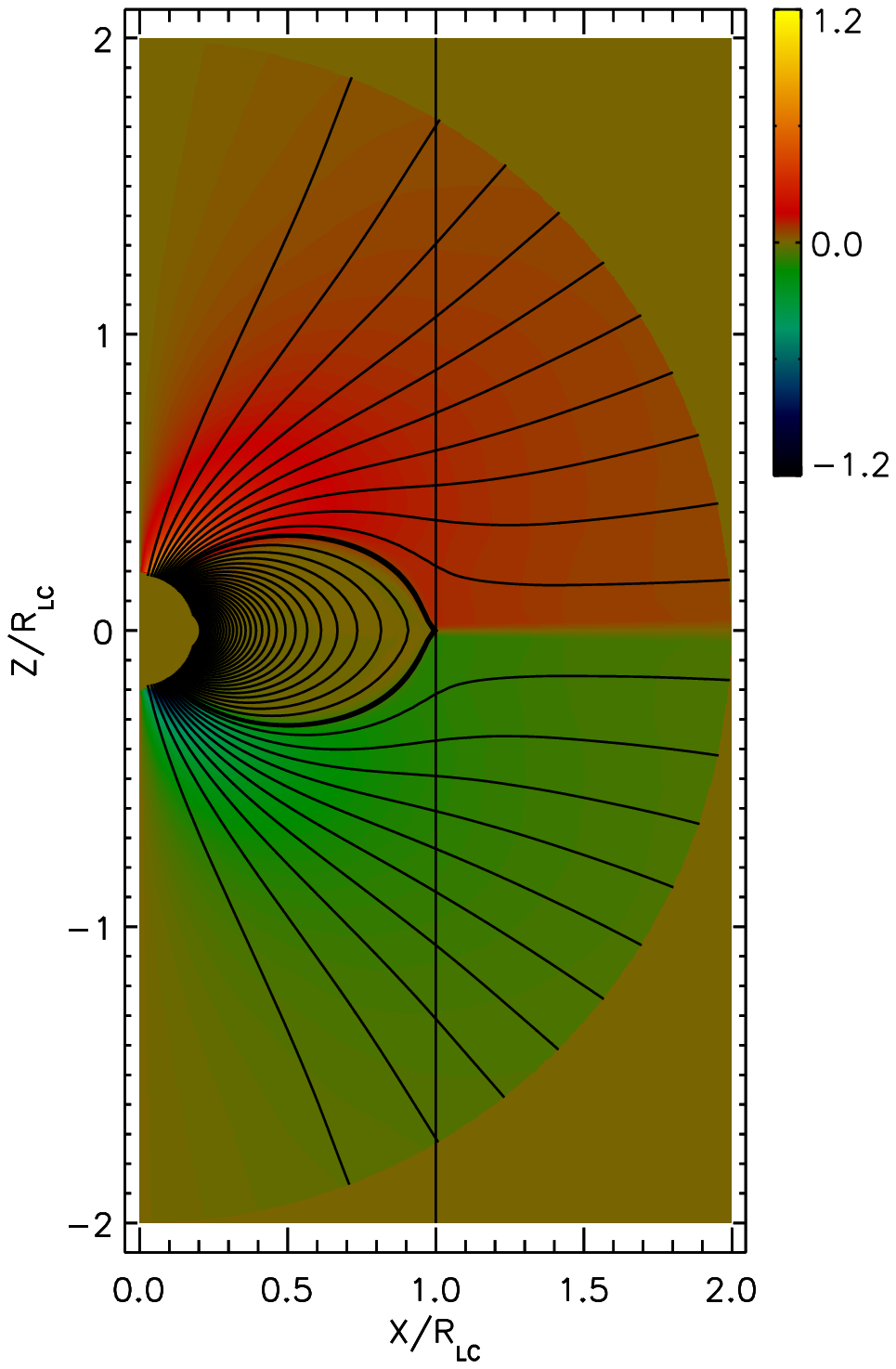}
\put(-270,400){a)}
\end{picture}
\hspace{1\unitlength}
\begin{picture}(280,400)(0,20)
\includegraphics[scale=.57]{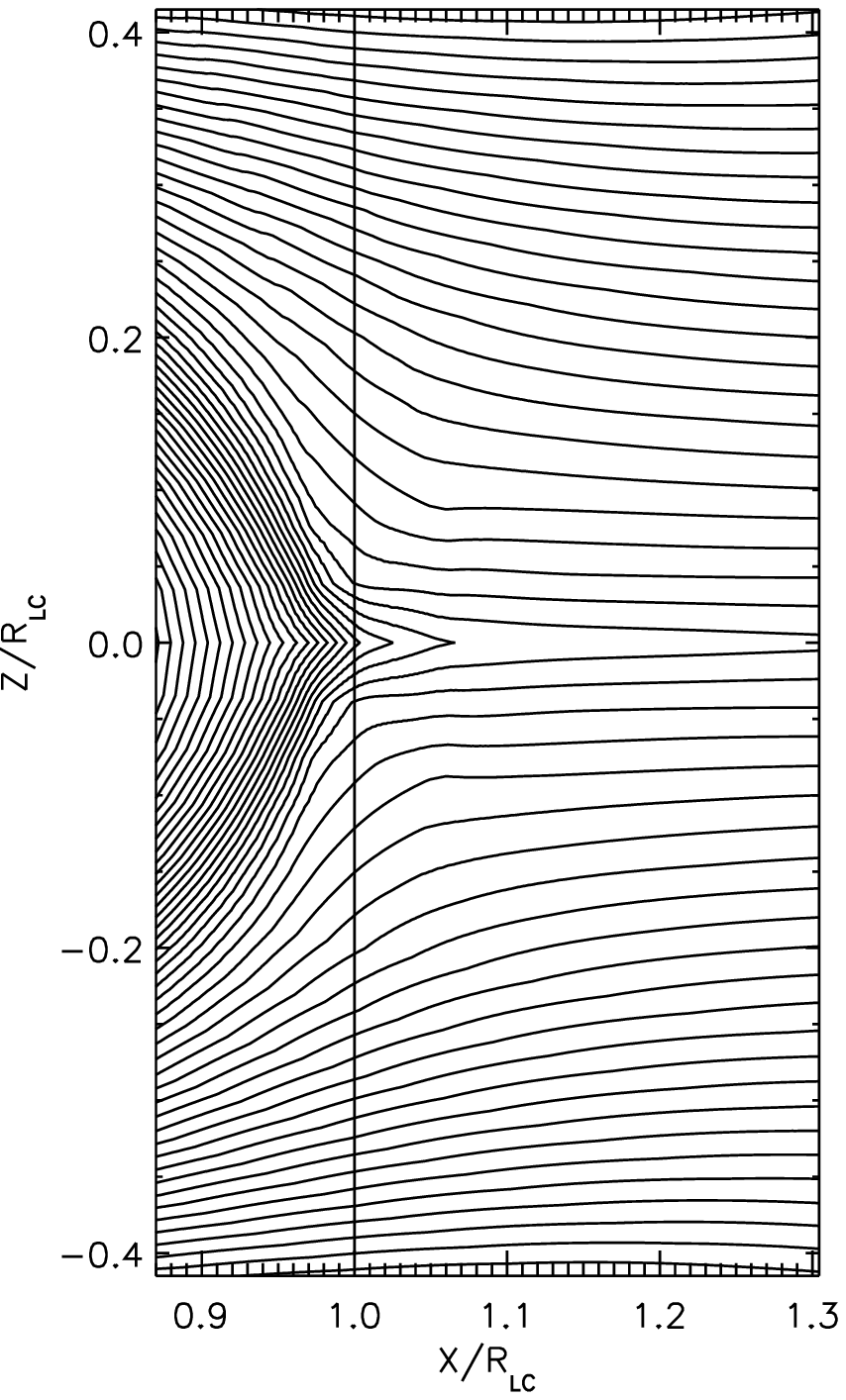}
\put(-280,400){b)}
\end{picture}
\hspace{1\unitlength}
\begin{picture}(280,400)(0,20)
\includegraphics[scale=.55]{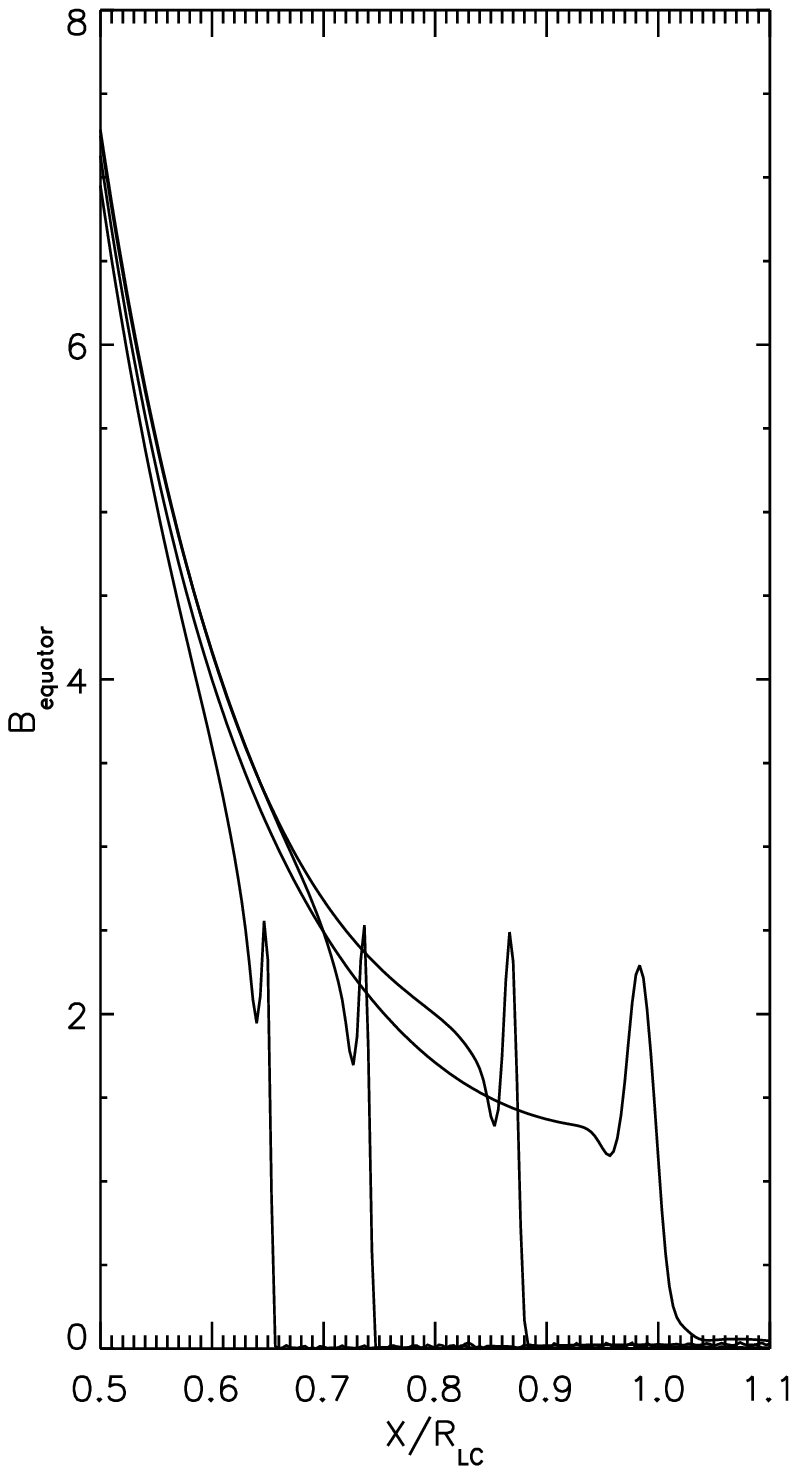}
\put(-260,400){c)}
\end{picture}
\hspace{1\unitlength}
\caption{
Aligned dipole magnetosphere: a) Poloidal fieldlines of steady state solution. Thick line is the fieldline that touches the light cylinder, which is marked by a vertical line. Color represents toroidal magnetic field, normalized by $\mu/(R_{LC}R_*)^{3/2}$; b)  Zoom in near the Y-point. Closed lines beyond the light cylinder are periodically thrown open by plasmoid emission; c) Magnitude of the magnetic field on the equator (in units of $\mu/R^3_{LC}$) during the evolution of the closed zone to the light cylinder. }
\label{figevol}
\end{figure*}

\section{Aligned pulsar magnetosphere}
Particle extraction from the stellar surface due to unipolar induction and subsequent pair-creation is likely to populate the pulsar magnetosphere with abundant nearly force-free plasma (Goldreich \& Julian 1969). The detailed structure of the magnetosphere had remained a puzzle for a long time due to the difficulty of finding an analytical solution. The first numerical solution to the time-independent set of force-free equations (the ``pulsar equation") was found by Contopoulos et al. (1999; hereafter, CKF), and later improved by Gruzinov (2005a). The CKF solution consists of a region of closed fieldlines extending to the light cylinder (``closed zone"), and an ``open zone" with asymptotically monopolar poloidal fieldlines. The requirement that the closed zone extend to the light cylinder was imposed from the outset, and was later relaxed by Goodwin et al. (2004), Contopoulos (2005) and Timokhin (2005), who find steady-state numerical solutions with closed zones of arbitrary extent within the light cylinder. This raises the question which solution would be chosen by a time-dependent evolution. Recently, Komissarov (2005) used a time-dependent force-free code and a relativistic MHD code for this problem. While his force-free code had converged to a solution with all closed fieldlines, presumably due to excessive numerical diffusion, the MHD code produced a solution remarkably similar to CKF. The force-free code of McKinney (2006b) also converged to the CKF result. Generally, we are in agreement with these results, however, intrinsic low dissipation nature of our code allows us to find features that were missing in prior studies. 

We discretize FFE equations in spherical coordinates on a uniform $1400\times 181$ $r-\theta$ grid with an absorbing zone to prevent reflections from the edge of the domain. The star has the radius $R_*=0.2R_{LC}$, and the light cylinder $R_{LC}\equiv c/\Omega_*$ is at 300 cells. The magnetic field is initialized to be a dipole $B_r=2 \mu \cos\theta/r^3$, $B_\theta=\mu \sin\theta/r^3$, $B_\phi=0$, where $\mu$ is the magnetic moment.
The electric field on the star is set to ${\bf E }=- {\Omega_*} \times {\bf r} \times {\bf B}/c$ to simulate a rotating conducting sphere, and the rotation is turned on as a step function at $t=0$. During the initial evolution a torsional Alfven wave is emitted from the surface (Spitkovsky 2004, 2005). The wave transports space charge, and the associated electric fields set fieldlines in rotation. Near the star cancellation of waves from two hemispheres causes a growing region with zero poloidal current on closed fieldlines. Further from the star the fieldlines get stretched, and a pulse propagates along the equator. In {\it ideal} FFE the fieldlines cannot break or reconnect and the solution tries to stretch the fieldlines so they close at infinity. In the process the magnetic field strength on the equator drops and becomes smaller than the electric field, necessitating non-ideal physics, and eventually hits zero. This happens around $0.6 R_{LC}$ after $1/4$ of stellar rotation. Beyond this point the solution forms a current sheet, supported by our resistivity prescription that keeps $E\leq B$. After 1 rotational period the magnetosphere forms the closed-open configuration, but with the Y-point at $0.6R_{LC}$. This is not the CKF solution, 
and in fact the configuration proceeds to expand the closed zone to $R_{LC}$ by {\it reconnection} of the recently opened fieldlines. The rate of this expansion depends on the resistivity prescription and is 20 rotation periods of the star for our lowest resistivity model, but can be accelerated to 5 rotation periods if we allow for $E$ to exceed $B$ in the current sheet by limiting the resistive current. 

Once the closed zone reaches $R_{LC}$, the solution becomes quasi-steady (Fig. 1a), and similar to CKF and McKinney (2006b). The fieldlines that reconnect outside $R_{LC}$ (Fig. 1b)  break through emission of small plasmoids every 5 periods, dependent on resistivity, and the Y-point oscillates around $(0.95 \pm 0.05) R_{LC}$.  
There is a sharp rise in the magnitude of the magnetic field on the equator at the Y-point (Gruzinov 2005a), followed by a jump by a factor of $10^2$ to nearly 0 in the current sheet (fig. 1c, corresponding jump is a factor of 10 in fig. 3 of McKinney (2006b), suggesting larger diffusivity). 
The evolution towards the light cylinder after the initial formation of the current sheet takes much longer than the light cylinder crossing time, and the magnetosphere goes through a sequence of quasi-static equilibria similar to the states in Timokhin (2005) (see fig. 1c). This evolution will be missed if code diffusivity is increased. The spindown energy loss 
decreases as the closed zone expands and saturates at $(1\pm 0.05) \mu^2 \Omega_*^4/c^3$, consistent with previous works. 
  \begin{figure*} 
  \vskip -.4in
\unitlength = 0.0011\textwidth
\hspace{40\unitlength}
\begin{picture}(400,400)(0,0)
\includegraphics[scale=.25]{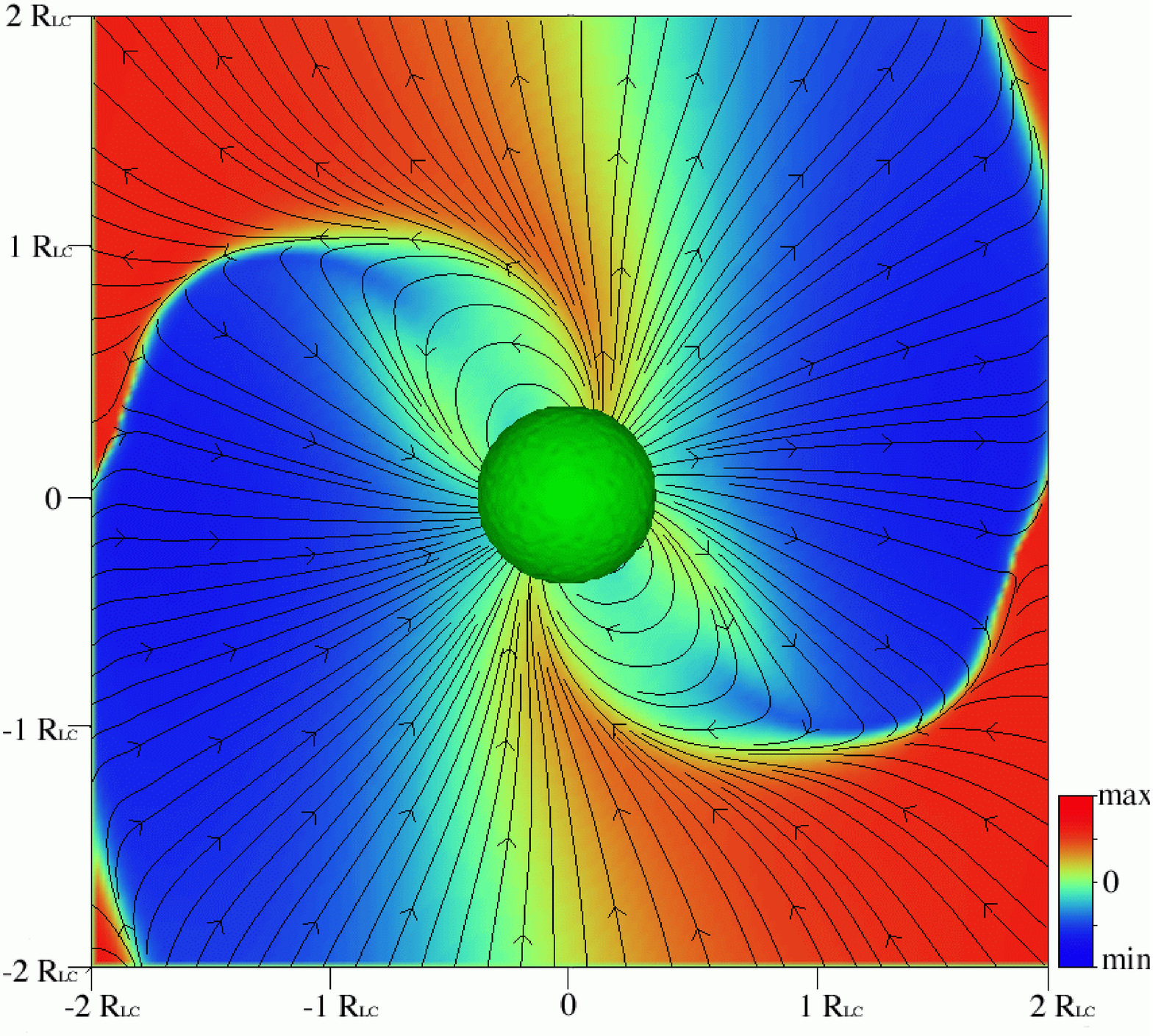}
\put(-390,300){a)}
\end{picture}
\hspace{15\unitlength}
\begin{picture}(400,400)(0,0)
\includegraphics[scale=.5]{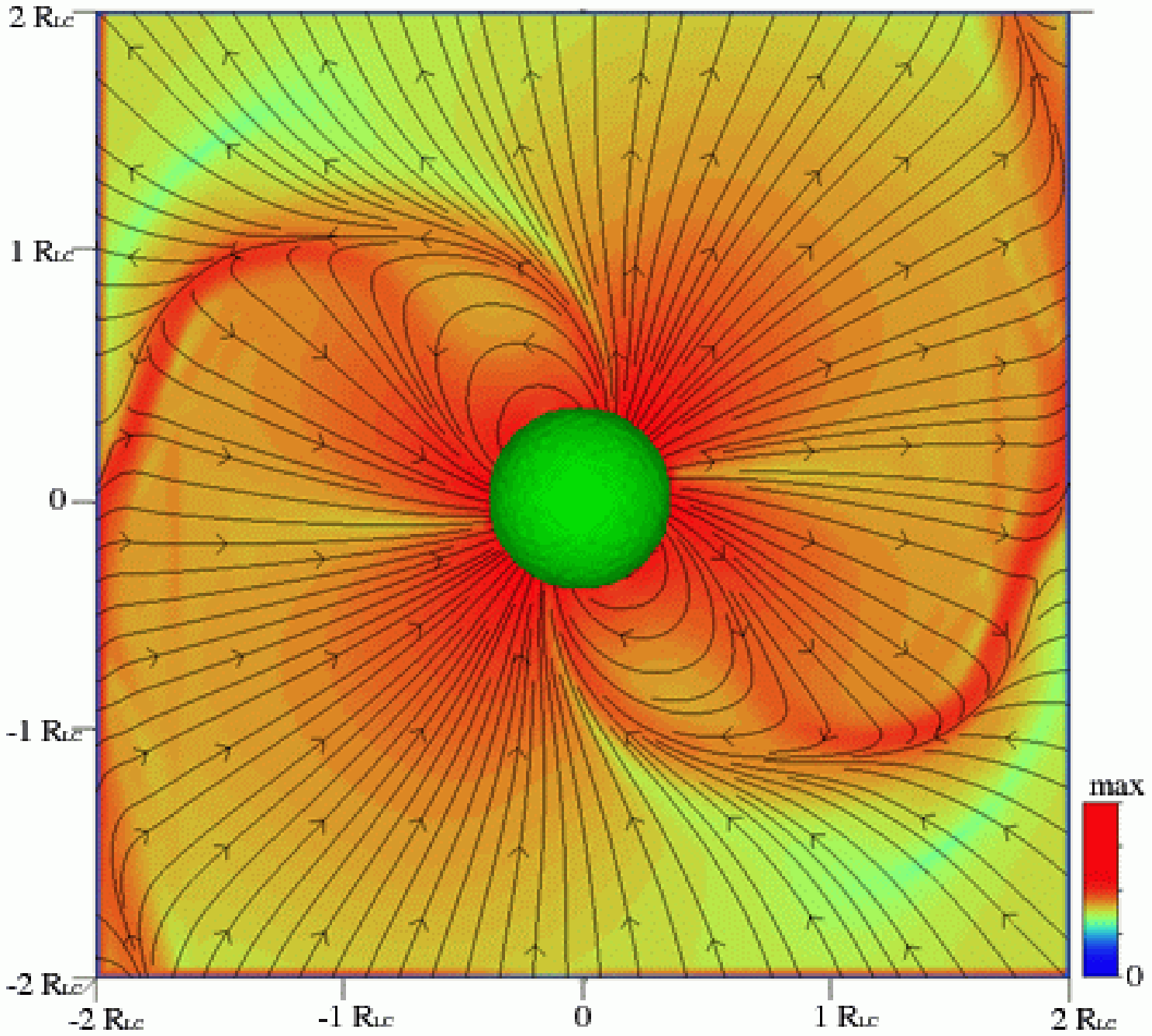} 
\put(-390,300){b)}
\end{picture}
\caption{
Oblique pulsar magnetosphere with magnetic inclination $\alpha=60^\circ$ in the corotating frame: a) Magnetic fieldlines in the ${\bf \mu}$-${\bf \Omega_*}$ plane. Color is the magnetic field perpendicular to the plane; b) same as a) but color represents absolute value of the total current $|{\bf \nabla} \times {\bf B}|$.}
\label{figobl}
\end{figure*}
  
\begin{figure*} 
\vskip -.4in
\unitlength = 0.0011\textwidth
\hspace{40\unitlength}
\begin{picture}(400,400)(0,0)
\includegraphics[scale=.5]{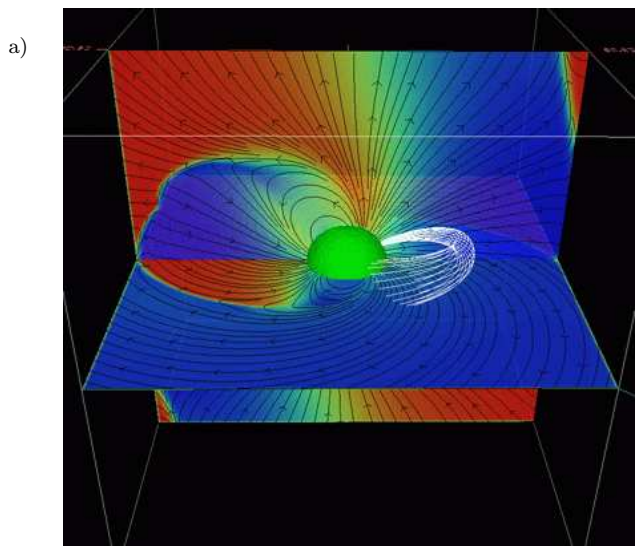}
\put(-420,330){a)}
\end{picture}
\hspace{-10\unitlength}
\hskip -.15in
\begin{picture}(400,400)(0,10)
\includegraphics[scale=.6]{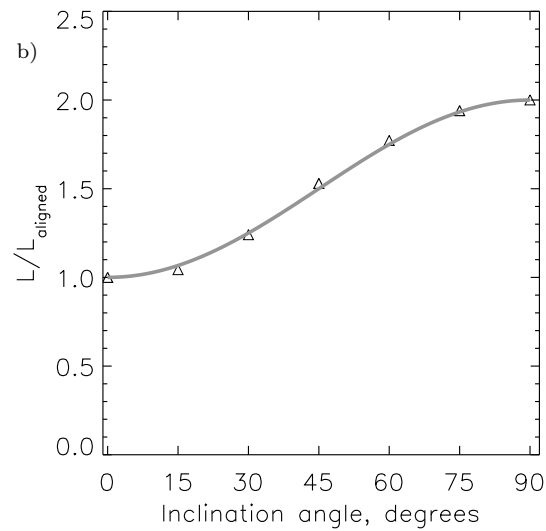} 
\put(-380,330){b)}
\end{picture}
\caption{
a) Slices through the $60^\circ$ magnetosphere. Shown are fieldlines in the horizontal and vertical plane, color on the vertical plane is the perpendicular field, on the horizontal plane -- toroidal field. Sample 3D flux tube is traced in white. b) Spindown luminosity in units of aligned force-free luminosity as a function of inclination. Triangles represent simulation data, and the line is a fit with eq. (3). }
\label{figobl3d}
\end{figure*}

\section{Oblique pulsar magnetosphere}
Real pulsars necessarily have nonzero angle between the magnetic and rotational axes (otherwise they would not pulse!), so understanding the structure of oblique pulsar magnetosphere is most interesting. Due to the lack of axial symmetry this problem has to be solved in full 3D. Encouraged by the convergence of our method to known solutions for the aligned rotator, we extended our code to three dimensions.  
Below we discuss gross properties of the field structure and energy loss.

We use a Cartesian grid which avoids coordinate singularities, but presents difficulties in imposing boundary conditions on a central sphere.  Rather than using stairstepping we force the fields to known values inside the star with a smoothing kernel (similar to Komissarov 2005). 
Inside the sphere the field is set to ${\bf{B}}_0(t)=(3\hat{\bf r}\vec{\bf{\mu}}(t)\cdot \hat{\bf r}-\vec{\bf{\mu}}(t))/r^3$, where $\vec{\bf{\mu}}(t)=\mu (\sin\alpha\cos\Omega_* t, \sin\alpha\sin\Omega_* t,\cos\alpha)$, and $\alpha$ is magnetic inclination angle. Electric field is forced to corotation values. We use a grid of $700^3$ cells with $R_*=15$ cells and $R_{LC}=40$ cells. Such unrealistic ratio $R_*/R_{LC}$ is needed to resolve both the star and the wavelength $2\pi R_{LC}$ on a limited grid. Without nonreflecting boundary  conditions we only evolved the magnetosphere for 2 turns of the star, but this seems to be enough to establish steady state solution in the corotating frame. 

Fig. 2 demonstrates the force-free magnetosphere of $\alpha=60^\circ$ rotator after 1.2 turns of the star. Although the magnetospheric structure is intrinsically 3D, some insight can be gained from considering the shape of the fieldlines in the corotating frame in the plane defined by $\bf{\mu}$ and $\bf{\Omega}_*$ vectors. In this plane the fieldlines are reminiscent of the CKF solution with a closed and open zone and a current sheet. The main difference is that the current sheet oscillates about the rotational equator in a wedge with the opening angle of $2\alpha$ and the wavelength of $2\pi R_{LC}$. The fieldlines in this plane become straight beyond the light cylinder, corresponding to the inclined split-monopole solution found by Bogovalov (1999). For comparison, the fieldlines of a rotating dipole in vacuum look dipolar at all radii in $\mu$-$\Omega_*$ plane, with no open zone or current sheet. In force-free solution the current sheet starts at the intersection of the light cylinder with the closed zone (fig. 2b) even in the oblique case. In the perpendicular plane the field is increasingly toroidal, reversing the sign in the current sheet (fig. 3a). 

An important question is the electromagnetic luminosity of the inclined rotator, since currently the vacuum formula $L_{\rm vac}=2/3\mu^2 \Omega_*^4/c^3 \sin^2 \alpha$ is commonly used to infer the magnetic field of pulsars. Our 3D solution allows a direct measurement of this quantity. We measure the Poynting flux integrated over spheres of different radius. After about 1 turn the solution settles to a constant energy flux, which we plot in fig. 3b as a function of magnetic inclination. We find that the  formula 
\begin{equation}
L_{\rm pulsar}=k_1 {\mu^2 \Omega_*^4\over c^3} (1+k_2 \sin^2\alpha), \label{sd}
\end{equation}
with coefficients  $k_1 = 1\pm 0.05$ and $k_2 = 1\pm 0.1$ gives a very good fit to the oblique spindown for all inclinations. This is roughly consistent with the estimate by Gruzinov (2005b). 
The inferred magnetic field at the magnetic equator of a standard neutron star is then $B_*=2.6\times 10^{19} (P \dot{P})^{1/2} (1+\sin^2\alpha)^{-1/2}$G which can be upto $1.7$ times smaller than the estimate from the vacuum formula. 
The fact that oblique rotators can lose upto two times more power and evolve faster in $P-\dot{P}$ space than aligned rotators reinforces the prediction that near the death line there should be an excess of pulsars with smaller inclination angles (Contopoulos \& Spitkovsky 2005). 

We have developed a numerical method for evolving time-dependent force-free MHD equations and have applied it to solving a dynamic pulsar magnetosphere. We reproduce the closed-open configuration of CKF for the aligned rotator, and extend the solution to the general oblique magnetic geometry, calculating the spindown rate as a function of inclination. We find that the shape of the magnetosphere is controlled by the physics of the Y-point. 
In order for the closed zone to reach the light cylinder reconnection must occur near the Y-point in the magnetosphere. If this reconnection is for some reason impeded, the magnetospheric extent may be smaller than the light cylinder. Reconnection of the open fieldlines must also happen when the pulsar spins down and the light cylinder recedes. If this reconnection is inefficient due to, for instance, intermittent plasmoid ejection or inertial loading, the Y-point may lag the light cylinder and lead to braking index less than 3 (Contopoulos \& Spitkovsky 2005). Y-point dissipation also leads to time-dependent phenomena, which when communicated to the inner magnetosphere, may leave an imprint on the radio emission (e.g. drifting subpulse phenomena). Such behavior can be modeled in our code with different prescriptions for resistivity, and further research into current sheet physics is needed. Availability of self-consistent 3D magnetospheric solutions opens the way to the development of quantitative models of phenomena in the magnetospheres of pulsars, magnetars and accretion disks. 
 

\end{document}